\documentclass[preprint2]{aastex}

\newcommand{\ang}{\AA}
\newcommand{\teff}{\ensuremath{T_{\rm eff}}}
\newcommand{\logg}{\ensuremath{\log (g)}}

\newcommand{\subsun}{\mbox{$_{\normalsize\odot}$}}

\begin{document}

\title{Effective Temperatures of late L~dwarfs
       and the onset of Methane signatures}

\author{Andreas Schweitzer}
\affil{Department of Physics and Astronomy \& Center for Simulational Physics, University of Georgia, Athens, GA 30602-2451 \email{andy@physast.uga.edu}}
    
\author{John E. Gizis\altaffilmark{1,2}}
\affil{Department of Physics and Astronomy,
University of Delaware,
Newark DE 19716\email{gizis@udel.edu}} 

\altaffiltext{1}{Previous address : Infrared Processing and Analysis Center, 100-22,
  California Institute of Technology, 
  Pasadena, CA 91125}

\author{Peter H. Hauschildt}
\affil{Department of Physics and Astronomy \& Center for Simulational Physics, University of Georgia, Athens, GA 30602-2451\email{yeti@hobbes.physast.uga.edu}}

\author{France Allard}
\affil{CRAL, Ecole Normale Superieure, 46 Alle d'Italie, Lyon, 69364, France \email{fallard@ens-lyon.fr}}

\author{Eric M. Howard\altaffilmark{2}}
\affil{Astronomy Department, University of Massachusetts, Amherst, MA 01003-4525\email{ehwd@kutath.astro.umass.edu}}

\and

\author{J. Davy Kirkpatrick}
\affil{Infrared Processing and Analysis Center, 100-22,
  California Institute of Technology, 
  Pasadena, CA 91125\email{davy@ipac.caltech.edu}}

\altaffiltext{2}{
Visiting Astronomer, Kitt Peak National Observatory, National Optical
Astronomy Observatory, which is
operated by the Association of Universities for Research in Astronomy,
Inc. (AURA) under cooperative
agreement with the National Science Foundation.
}

\begin{abstract}
We present a spectral analysis of a sample of late L~dwarfs. We use our latest
model atmospheres and synthetic spectra and optical and K band spectra to
determine effective temperatures.  We derive effective temperatures of 1400~K
to 1700~K for L8 to L6 dwarfs.  The analysis demonstrates that our recent
 models that rain out the formed dust completely are applicable to
 optical spectra of late L~dwarfs and that
more consistent models are needed for intermediate L~dwarfs and for infrared
spectra.  We compare the results for the effective temperatures with the
temperatures of the onset of Methane formation.  Our models predict Methane
absorption at 3.3\micron\ to occur at about 400~K higher temperatures than
Methane absorption at 2.2\micron. This is consistent with our data and previous
observations which show Methane absorption at 3.3\micron\ but not at
2.2\micron\ in late L~dwarfs.
\end{abstract}


\section{Introduction}

L~dwarfs \citep{martin97,martin98,martin99,zapatero99,kirk99,kirk2000} are dwarfs cooler than M~dwarfs and hotter than 
the so called Methane or T~dwarfs.
They are defined by showing decreasing or no TiO and VO absorption in the optical
and by not yet showing CH$_4$ in the near infrared.
There are two classification schemes for L~dwarfs, the one by
\citet{martin99} and the one by \citet{kirk99} which differ
especially for late L~dwarfs --- which are the topic of this
work. For the purpose of this paper
we adopt the scheme by \citet{kirk99}.
The transition from M~dwarfs to L~dwarfs is observationally very
smooth and theoretically well explained by gradually decreasing the
effective temperature.
The transition from L~dwarfs to T~dwarfs, however, is not as smooth.
The original definition of L~dwarfs deliberately left room between
L~dwarfs and cooler objects \citep{kirk99}.
First estimates on the effective temperatures of late L~dwarfs and on
the first known Methane dwarf Gl229B
did not clarify if there is a jump in effective temperature and if there
is a missing population of objects between L~dwarfs and Methane dwarfs.
Recent analyses of L~dwarfs and to some extend T~dwarfs can be found in e.g.
\citet{gl229ballard,marley96,tsuji96,tinney98,burrows2000,basri2000,pavlenko2000,lsandy}.
Those analyses range from optical to infrared analyses and they range
from measuring low to high resolution spectra.
However, the uncertainty about the effective temperatures and the transition
to the T~dwarfs remains.

One aspect of this transition is the onset of Methane absorption.
L~dwarfs do not show Methane absorption in the near infrared, however
Gl229B and even Jupiter do. Recently, \citet{noll2000} reported
the detection of Methane at 3.3\micron\ in late L~dwarfs which
seems to close the gap between L~dwarfs and Methane dwarfs.
And as discussed in \citet{noll2000} it even makes the naming
convention ``Methane dwarf'' ambiguous.

Regardless, the effective temperatures associated with late
L~dwarfs and the onset of Methane absorption is not clear, yet.
\citet{noll2000} estimated relatively high effective temperatures
of about 1800~K for late L~dwarfs, whereas other studies suggest
effective temperatures as low as 1300~K for the latest L~dwarfs \citep{kirk2000}.
Note that the effective temperature of Gl229B is estimated
to be 900--1000~K \citep{gl229ballard,marley96}, but that it is not the warmest
T~dwarf since earlier ones \citep[e.g.][]{leggett2000sloan,burgasser2000,burgasser2001}
are known to exist.
This work aims to clarify the effective temperatures
of late L~dwarfs and the onset of Methane absorption.

In the next two sections we describe the data and the models
used in this work, respectively.
In section 4 we present the analysis which consists of an optical
spectral analysis, an infrared spectral analysis and an absolute
flux analysis for the objects with known parallaxes.
We discuss and conclude this work in section 5.

\section{The data}

The optical spectra for the stars of this sample are taken from
\citet{kirk99} and \citet{kirk2000}. The K-band spectra used in this
work were obtained at the Kitt Peak
4-meter telescope using the CRSP spectrograph.
   The spectra were reduced using standard techniques.
   The atmospheric water absorption was removed using
   nearby A type stars.
See also \citet{gizis2001} for details of the observations.
The sample constitutes of all late ($\ge$ L6) L~dwarfs that have
both optical and K band spectra.
The criterion L6 was physically relatively arbitrary. It was a
rather technical criterion. We wanted to
comfortably include the spectral range used in \cite{noll2000}
and we wanted to have a spectral range for which we have valid models.
See \citet{cond-dusty,lhires}
and sections \ref{models} and \ref{summary} for
a discussion of the validity of the applied models.
Especially the respective sections of this paper will make this work
also a case study for the validity of the AMES-Cond models.

Some of the stars also
have known trigonometric parallaxes which are collected
and tabulated in \citet{kirk2000}.
All objects and their properties are listed in Table \ref{lowresfittab}.

\section{The models}
\label{models}

The models used in this work are the so called AMES-Cond and AMES-Dusty models described
in \citet{cond-dusty}. A detailed description can be found therein.
The key features relevant for this study are the dust treatment
and the latest line lists which are summarized in the next few
paragraphs.

In the AMES-Cond models the dust is treated in the limiting case
where the dust has formed in equilibrium and has rained out completely out
of the atmosphere.
This effectively removes the fraction of elements from the atmosphere that are bound
in dust grains; see \citet{cond-dusty} for a discussion.
The AMES-Dusty models treat the dust in the other limiting
case where the dust stays in the layer in which it forms and
does not rain out at all. This effectively introduces
a dust opacity due to the full dust content.
To calculate the dust content
we use the Gibbs free energy 
of formation from   \citet{sharp_huebner1990}
and calculate the equilibrium concentration of all dust species
available.

In earlier analyses, the models have been applied to low resolution optical and infrared
spectra of M and L dwarfs \citep{lsandy} and to optical low and high
resolution spectra of early L~dwarfs \citep{lhires}. Earlier models
have been applied to high resolution spectra of L~dwarfs as well
\citep{basri2000}. The results from these works indicate
that the assumption of complete rain out is valid for
late L~dwarfs. Therefore, we will primarily apply the AMES-Cond models
in our analysis of the late L~dwarfs.
However, to further investigate the condition
of complete rain out we also used infrared spectra of AMES-Dusty
models in sections \ref{ir} and \ref{methane} as comparison.

The important opacity sources included in this work are the
TiO\citep{ames-tio}, FeH\citep{phillips93}, CrH\citep[Freedman, private comm., see
also][]{cond-dusty}, H$_2$O\citep{ames-water-new} and CH$_4$\citep{geisa92} line lists.
This means that all important opacity sources for (late) L dwarfs are treated with
line lists and no longer with JOLA (Just Overlapping  Line Approximation) opacities.
However, we note that there are still uncertainties in the
data for the H$_2$O line list \citep{allard2000}.

The spectra used for this analysis are calculated at a resolution of 2~\ang.
They are based on fully converged model structures that include
depth depended line profiles and include all the opacities
for all iterations \citep[see e.g.][for details
of the calculations]{NGhot,cond-dusty}.

\section{Analysis}
\subsection{Optical spectra}

We used a  $\chi^2$ fitting technique to determine the
best fitting model spectra.
It is described in \citet{lsandy} and has also been used in
\citet{lhires}.
As input we used a grid of AMES-Cond model spectra from
\teff=1000~K to \teff=1700~K in steps of 100~K and
from \logg=4.0 to \logg=6.0 in steps of 0.5 to
determine which model results in the smallest $\chi^2$
value.
Since the observed spectra have a resolution of
significantly less than R=1000 the standard resolution of 2~\ang\ of our model spectra
is sufficient for this analysis. However, the resolution of
the observations is not
high enough to analyze individual atomic lines.
Therefore, the $\chi^2$ method used in this work
measures the goodness of the fit to the overall spectrum and weighting
individual lines would not be allowed by the quality of the
available data.
We employ error bars for the measured values of 100~K in \teff and 0.5~dex in \logg.
This reflects primarily the grid spacing.
However, the three best fitting models do lie
within these error bars.
These error bars do not take into account any
systematic errors due to uncertainties of
the models.

\begin{figure}
\plotone{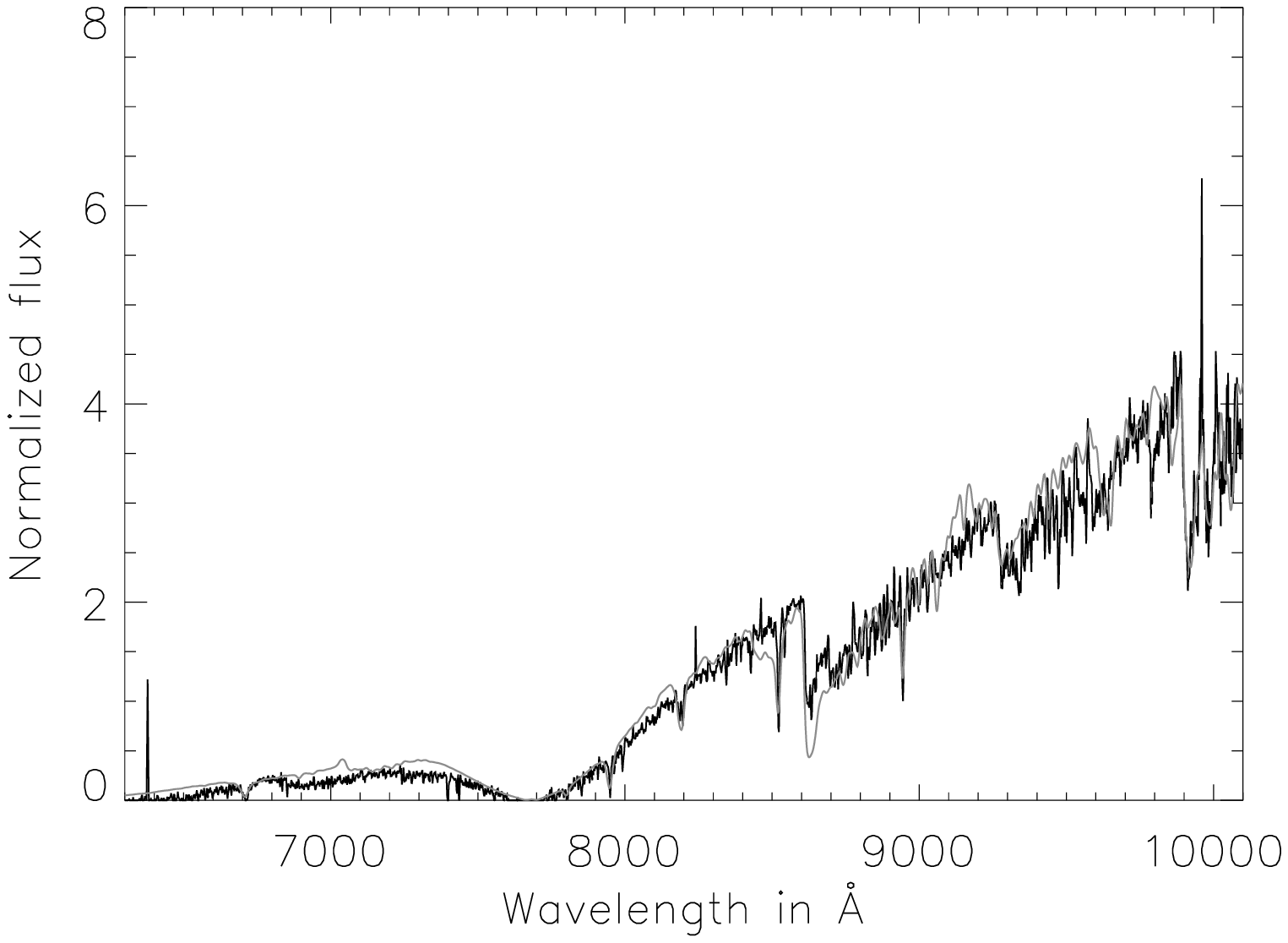}
\caption{\label{opt0850a}
Fits (grey line) to 2M0850+105 (dark line).
See Tab. \ref{lowresfittab} for parameters.
Telluric features have not been removed.
See \citet{kirk99} for exact locations of telluric bands.
}
\end{figure}

\begin{figure}
\plotone{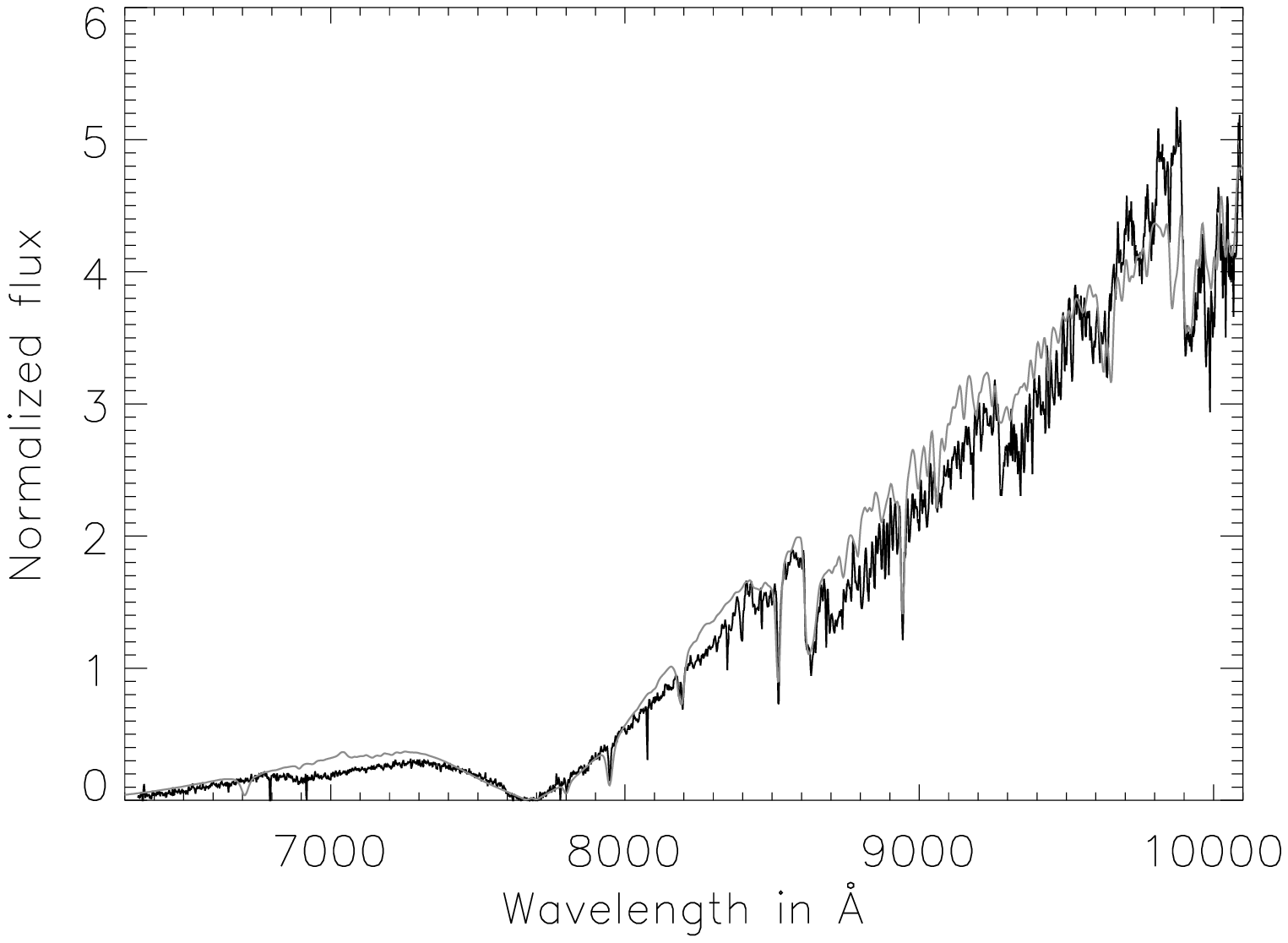}
\caption{\label{opt0920}
Fits (grey line) to 2M0920+351 (dark line).
See Tab. \ref{lowresfittab} for parameters.
Telluric features have not been removed.
See \citet{kirk99} for exact locations of telluric bands.
}
\end{figure}

\begin{figure}
\plotone{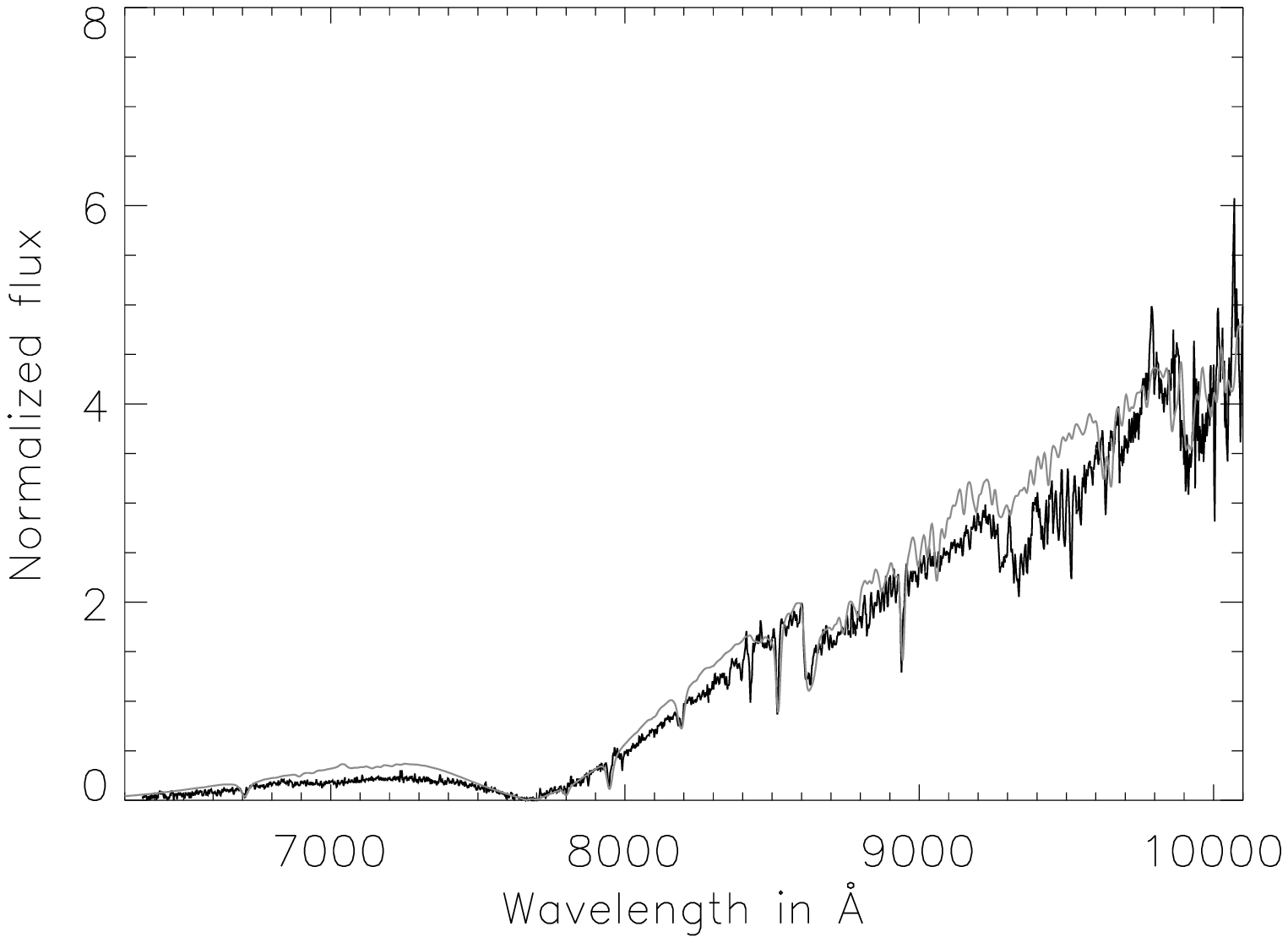}
\caption{\label{opt0825}
Fits (grey line) to 2M0825+2115 (dark line).
See Tab. \ref{lowresfittab} for parameters.
Telluric features have not been removed.
See \citet{kirk99} for exact locations of telluric bands.
}
\end{figure}

\begin{figure}
\plotone{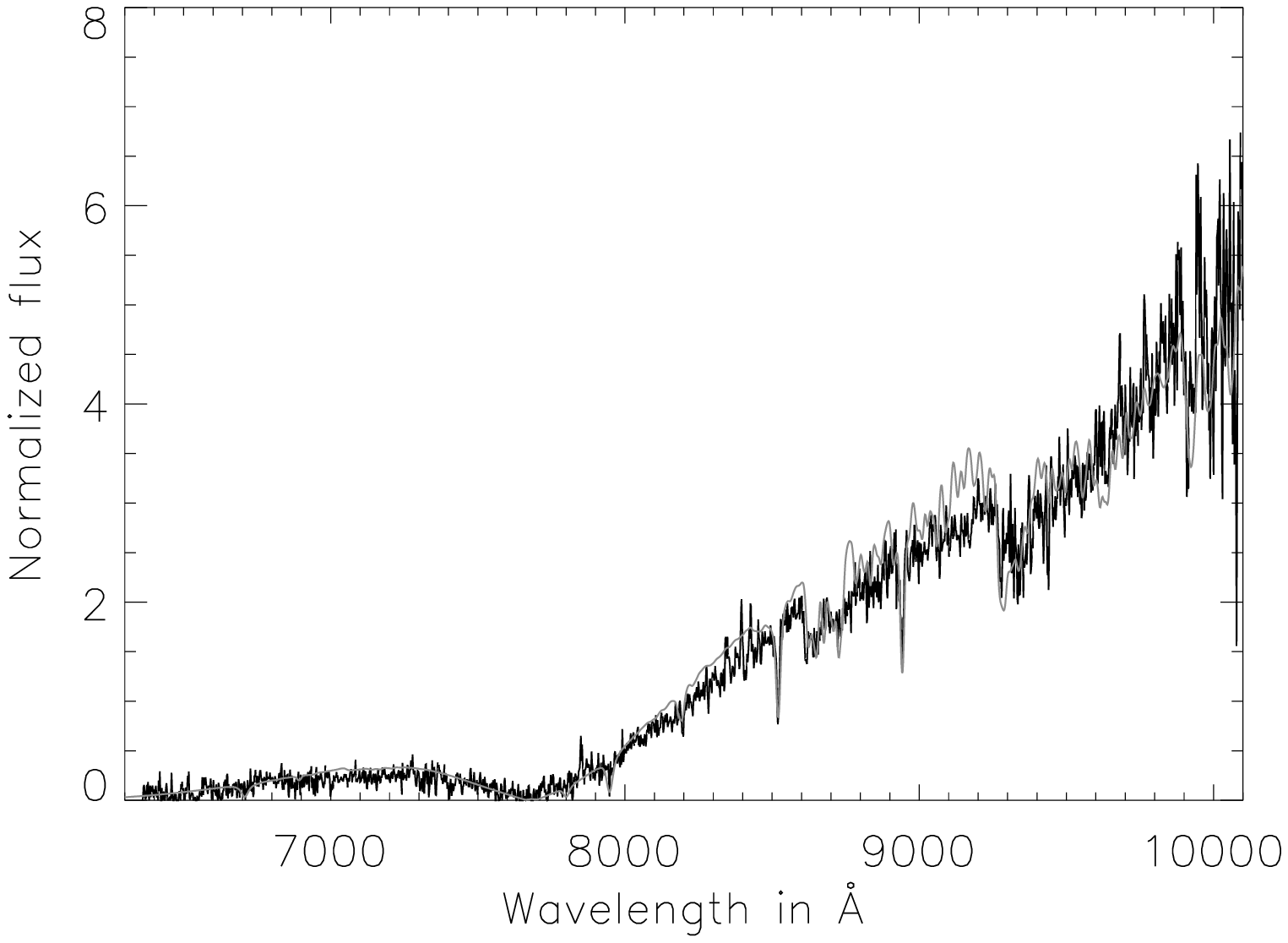}
\caption{\label{opt0929b}
Fits (grey line) to 2M0929+342 (dark line).
See Tab. \ref{lowresfittab} for parameters.
Telluric features have not been removed.
See \citet{kirk99} for exact locations of telluric bands.
}
\end{figure}

\begin{figure}
\plotone{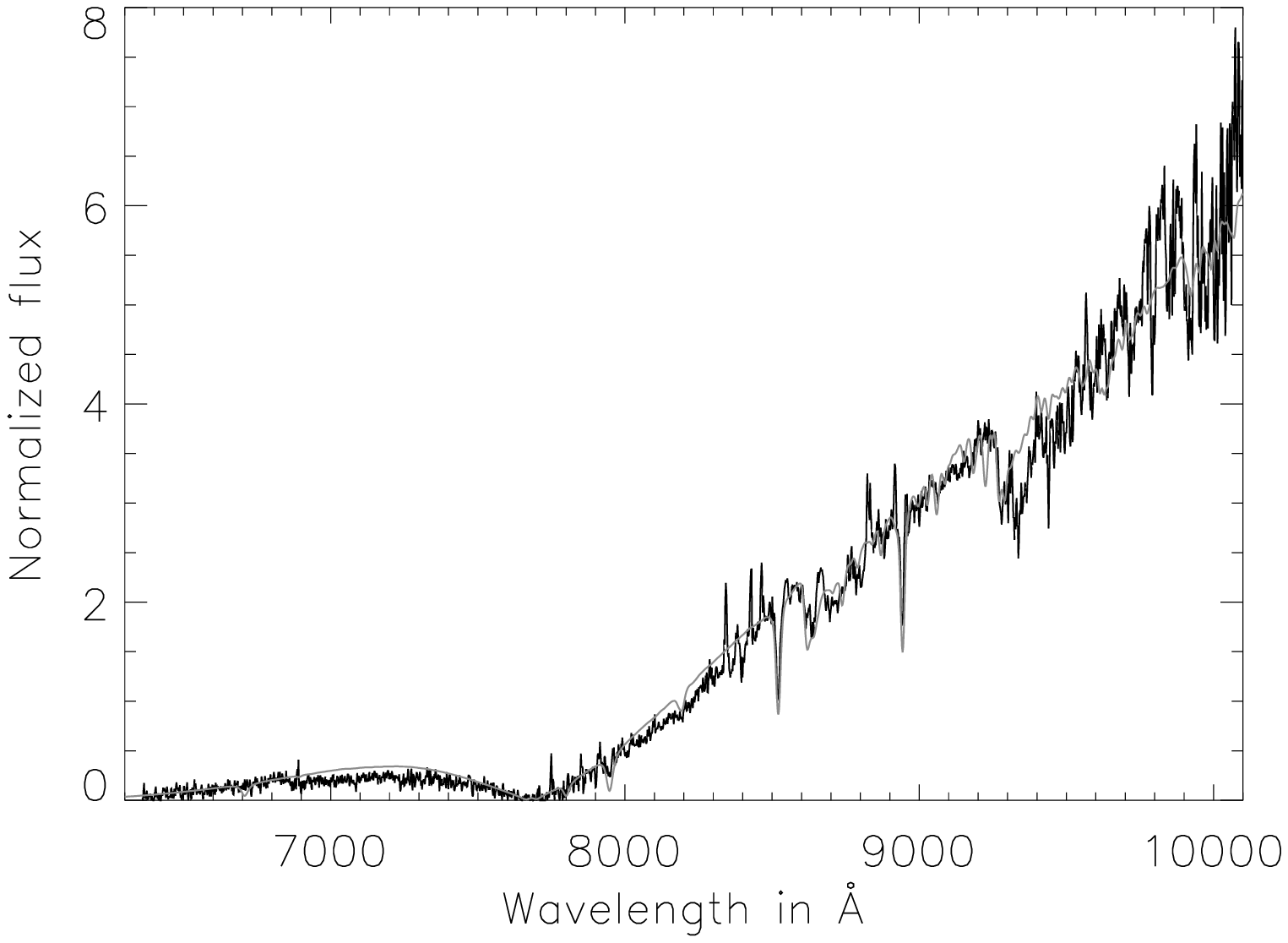}
\caption{\label{opt0328c}
Fits (grey line) to 2M0328+230 (dark line).
See Tab. \ref{lowresfittab} for parameters.
Telluric features have not been removed.
See \citet{kirk99} for exact locations of telluric bands.
}
\end{figure}

\begin{figure}
\plotone{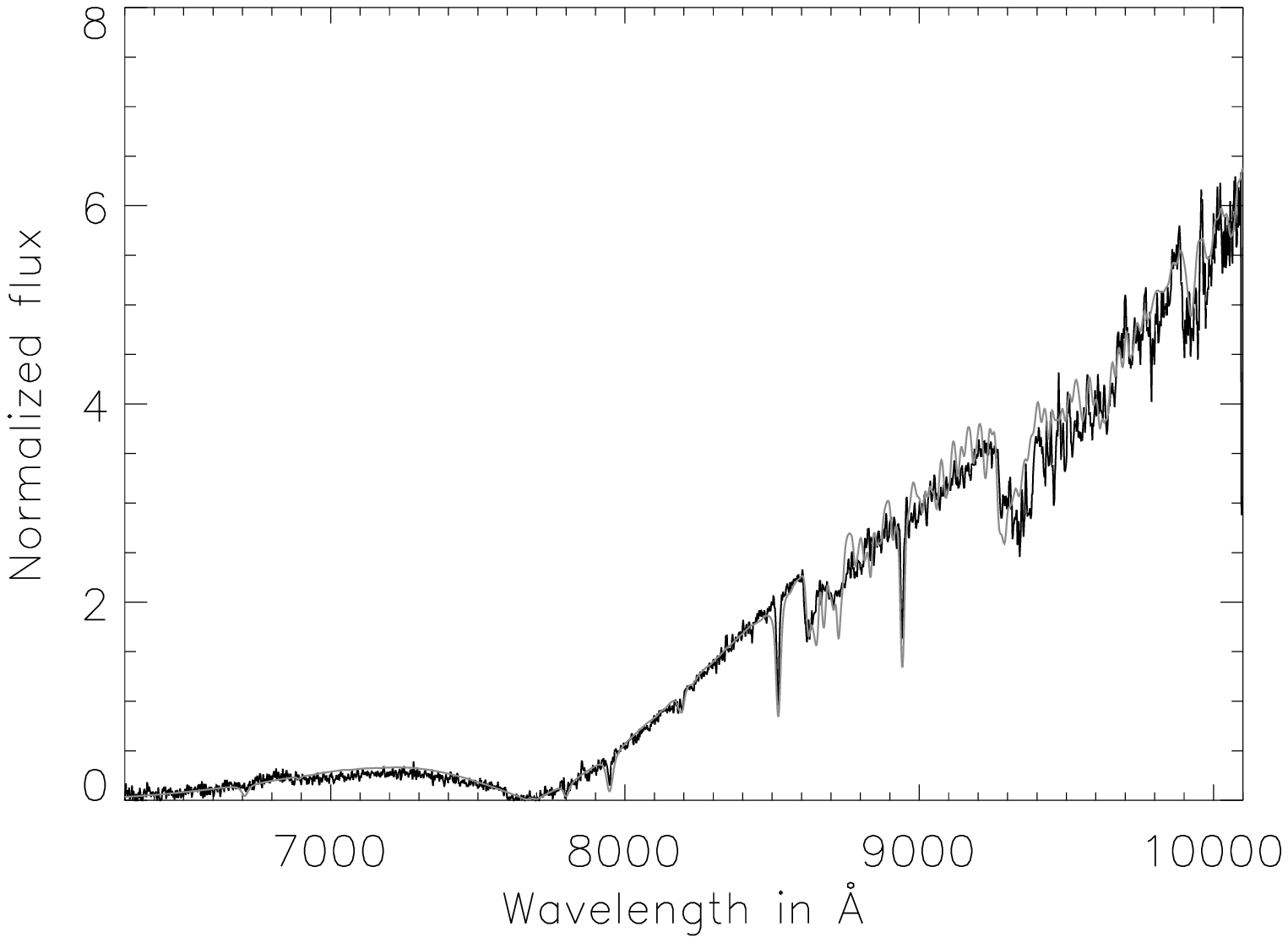}
\caption{\label{opt1632b}
Fits (grey line) to 2M1632+190 (dark line).
See Tab. \ref{lowresfittab} for parameters.
Telluric features have not been removed.
See \citet{kirk99} for exact locations of telluric bands.
}
\end{figure}

\begin{figure}
\plotone{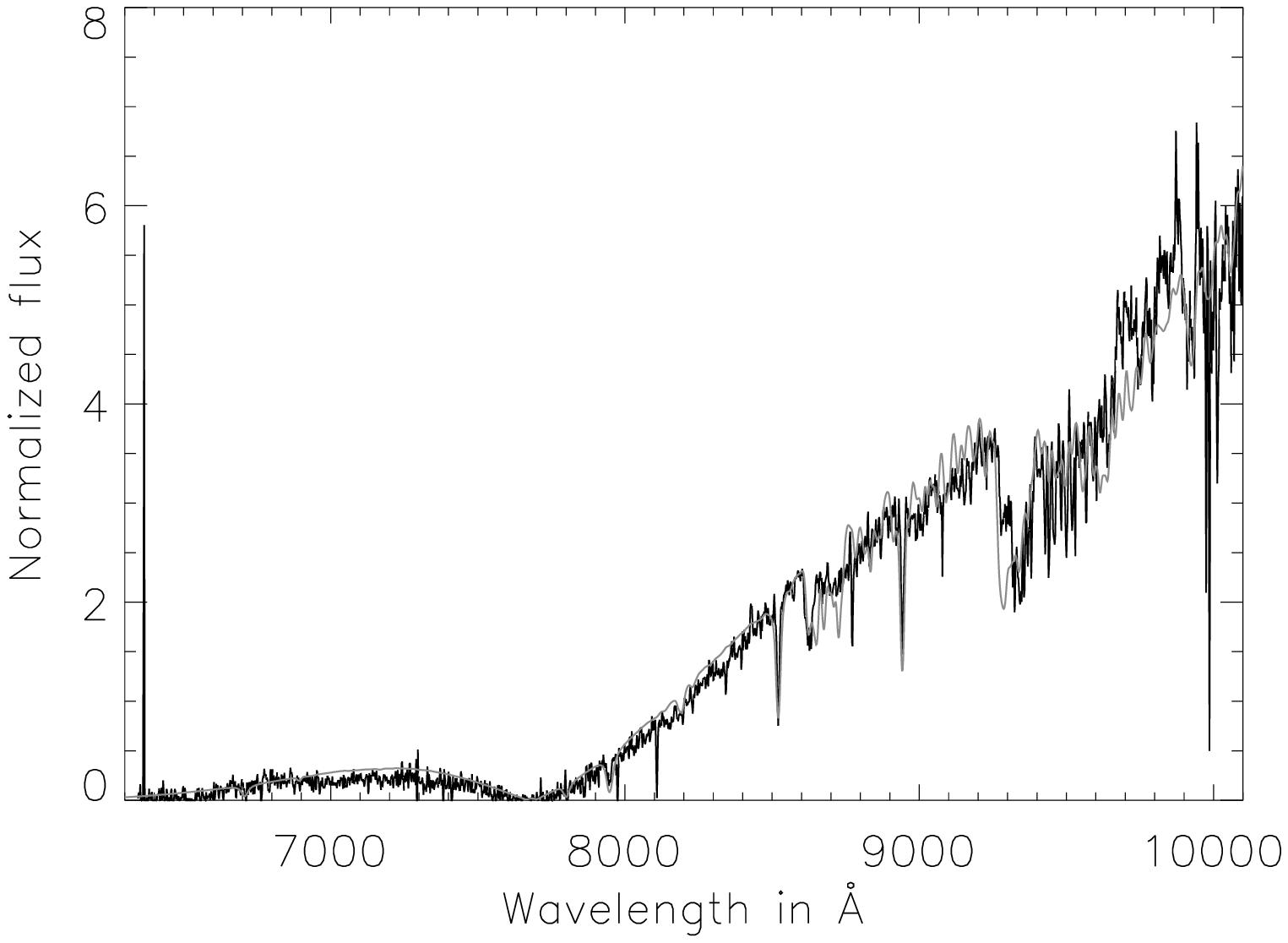}
\caption{\label{opt1523b}
Fits (grey line) to 2M1523+301 (dark line).
See Tab. \ref{lowresfittab} for parameters.
Telluric features have not been removed.
See \citet{kirk99} for exact locations of telluric bands.
}
\end{figure}

The best fits to the optical spectra are shown
in Figs.\ref{opt0850a} to \ref{opt1523b} and the
parameters for the best fitting models are summarized in
Tab.\ref{lowresfittab}.
As can be seen the models fit very well to the observations.
The FeH and CrH bands at 8600~\ang\ and the H$_2$O  bands at 9300~\ang\ are well reproduced as well
as the overall shape which is determined by the \ion{K}{1}
doublet at 7600~\ang\ and the Na~D doublet.
Also the \ion{Rb}{1} and \ion{Cs}{1} are reproduced within the limits
of the resolution.

The determined effective temperatures correlate with spectral type,
as expected. The results might suggest a jump in effective temperature
from L7 to L8 by 300K. However, this sample is too small to
draw such conclusions, especially within the error bars
of the fitting process.
The determined gravities are \logg=5.0 or larger which means
the objects are well contracted and not young anymore.
This is expected for a field sample like this 2MASS sample
and very similar to the results for \logg\ obtained in \citet{lhires}.

\subsection{Infrared spectra}
\label{ir}

\begin{figure}
\plotone{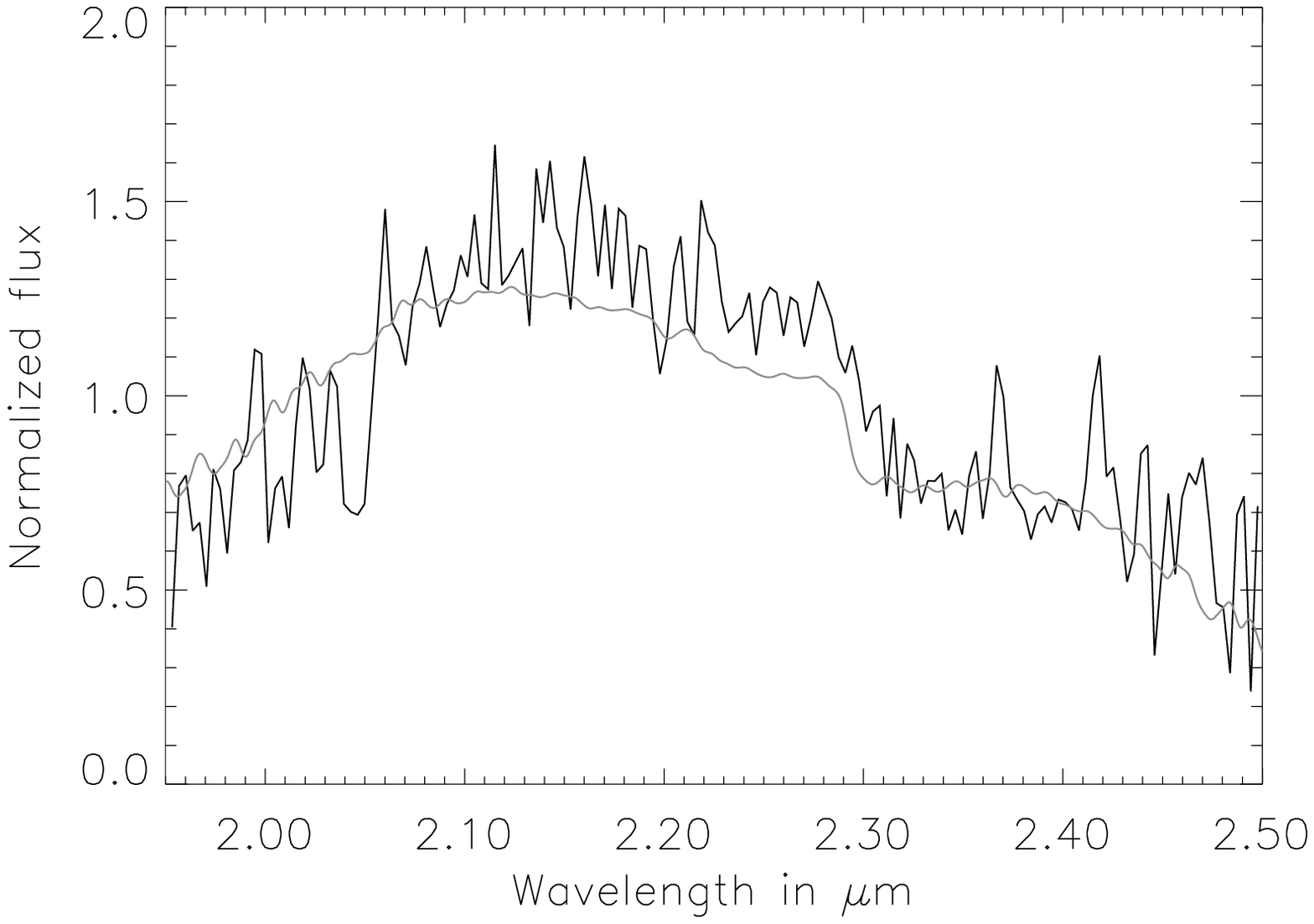}
\caption{\label{ir0850}
Fits (grey line) to 2M0850+105 (dark line).
See Tab. \ref{lowresfittab} for parameters.
}
\end{figure}

\begin{figure}
\plotone{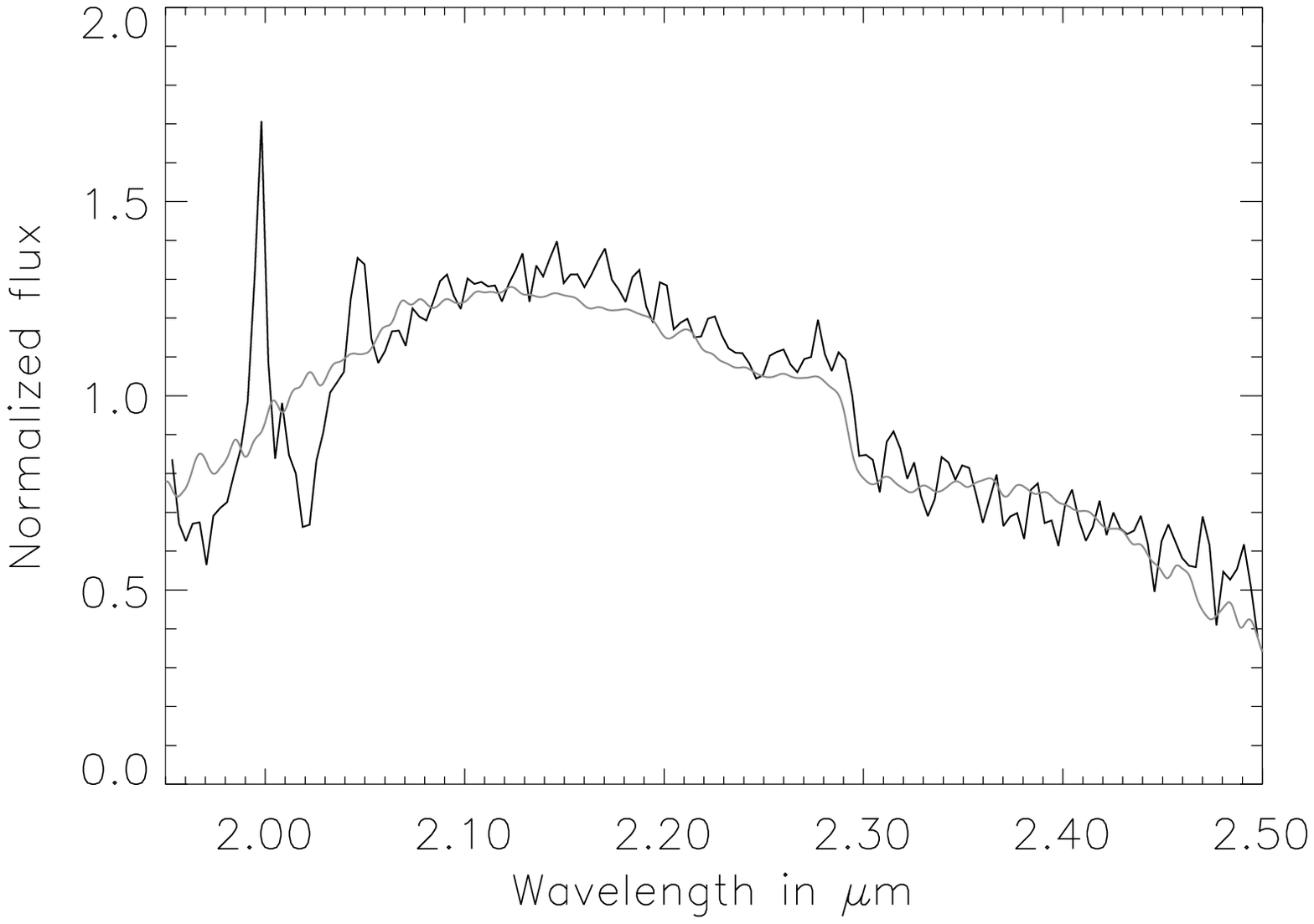}
\caption{\label{ir0920}
Fits (grey line) to 2M0920+351 (dark line).
See Tab. \ref{lowresfittab} for parameters.
}
\end{figure}

\begin{figure}
\plotone{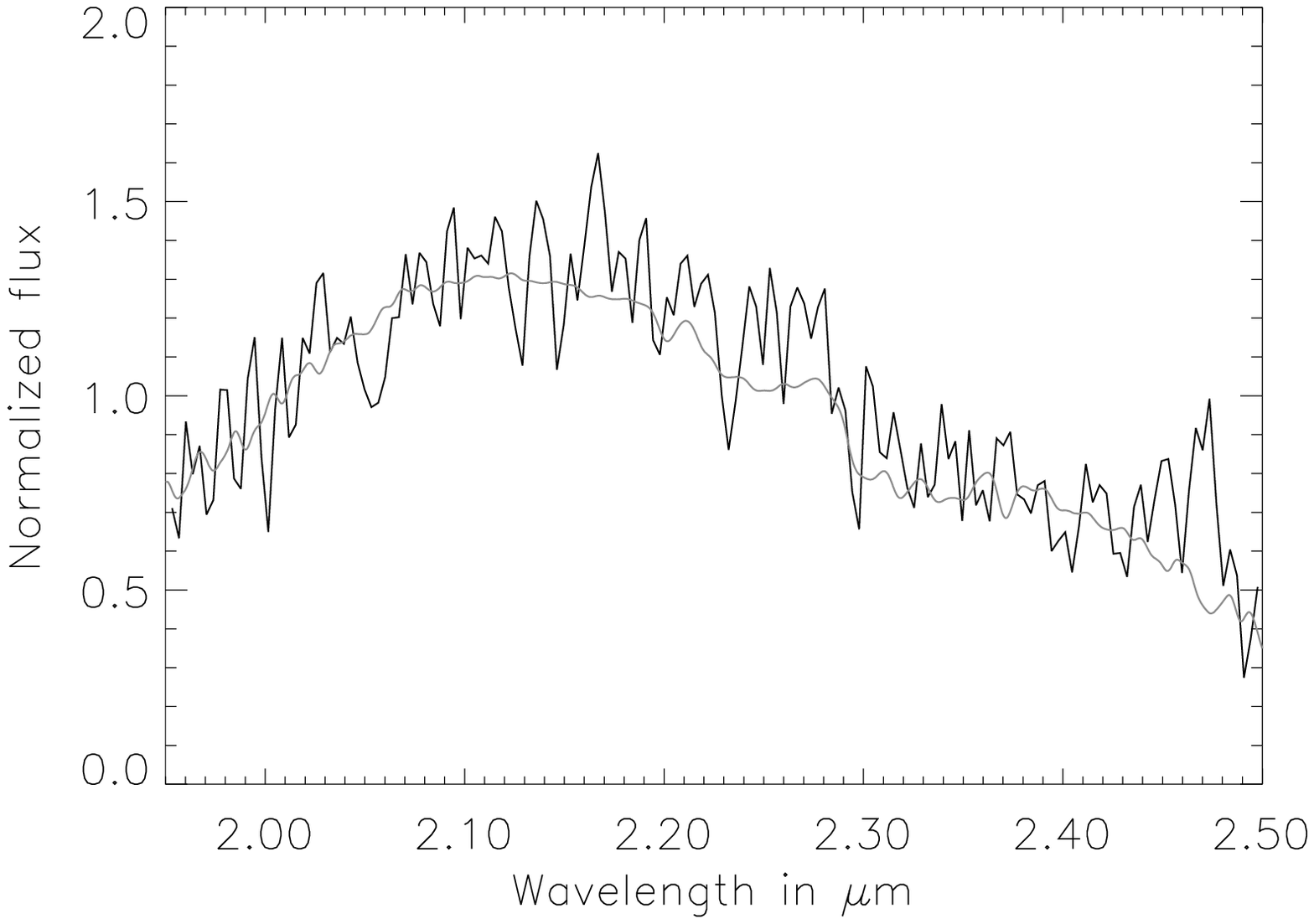}
\caption{\label{ir0825}
Fits (grey line) to 2M0825+2115 (dark line).
See Tab. \ref{lowresfittab} for parameters.
}
\end{figure}

\begin{figure}
\plotone{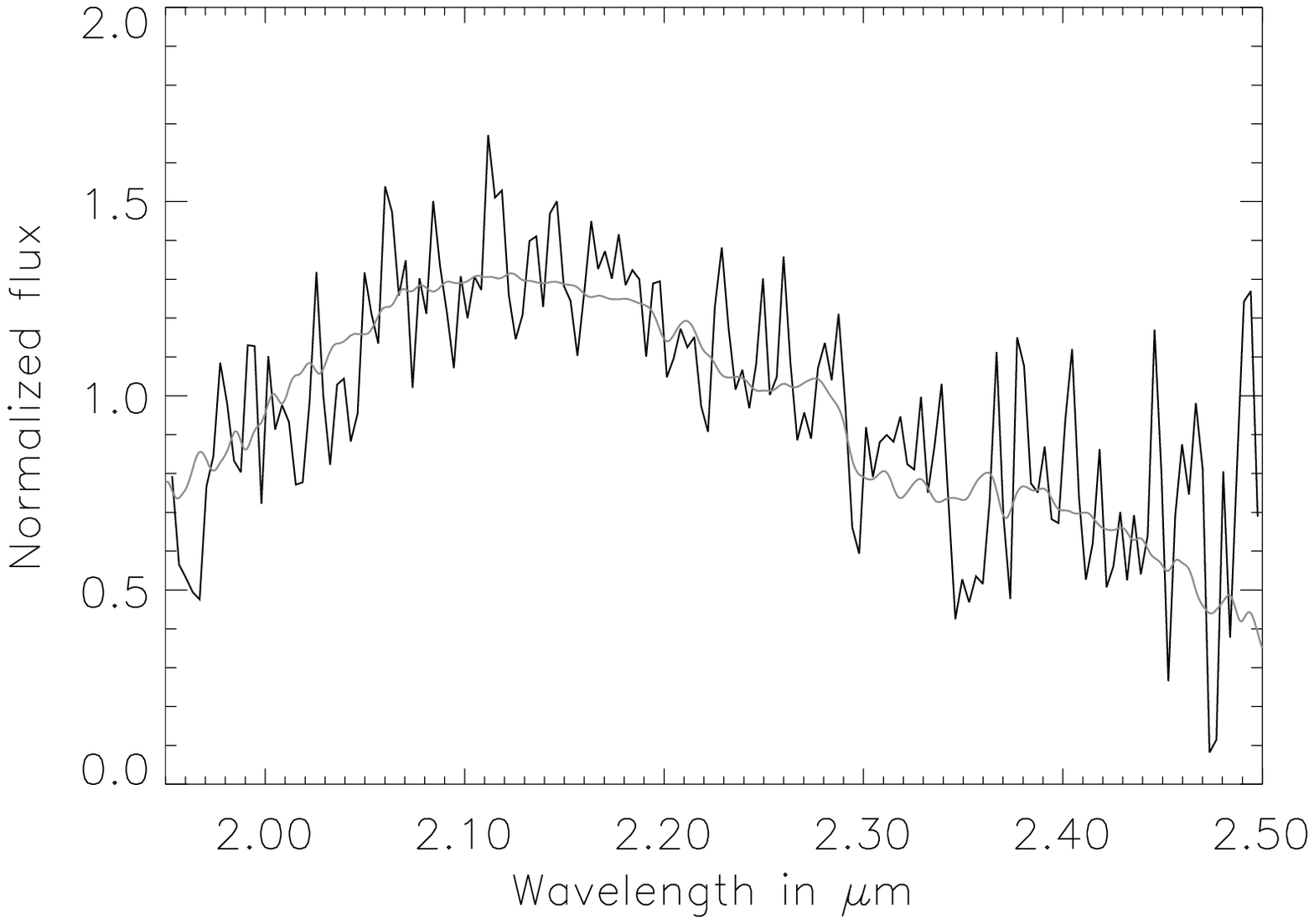}
\caption{\label{ir0929}
Fits (grey line) to 2M0929+342 (dark line).
See Tab. \ref{lowresfittab} for parameters.
}
\end{figure}

\begin{figure}
\plotone{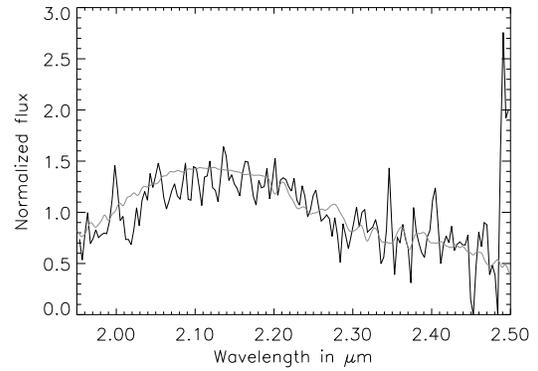}
\caption{\label{ir0328}
Fits (grey line) to 2M0328+230 (dark line).
See Tab. \ref{lowresfittab} for parameters.
}
\end{figure}

\begin{figure}
\plotone{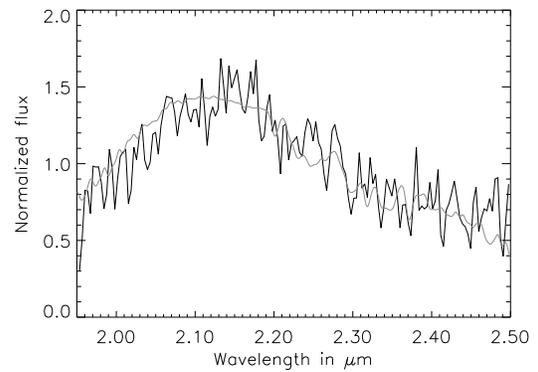}
\caption{\label{ir1632}
Fits (grey line) to 2M1632+190 (dark line).
See Tab. \ref{lowresfittab} for parameters.
}
\end{figure}

\begin{figure}
\plotone{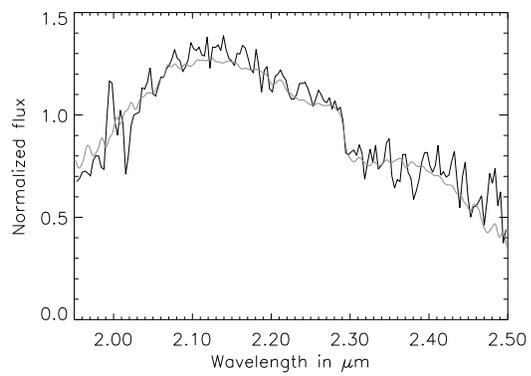}
\caption{\label{ir1523}
Fits (grey line) to 2M1523+301 (dark line).
See Tab. \ref{lowresfittab} for parameters.
}
\end{figure}

We performed the same type of $\chi^2$ fitting technique
for the infrared spectra as we did for the optical spectra.
The model grid for the infrared analysis had to encompass effective temperatures from
1300~K to 1900~K to obtain good fits.
The resulting fits are shown in Figs. \ref{ir0850} to \ref{ir1523} and the best
fitting parameters are listed in Tab. \ref{lowresfittab}.
For the K~band spectra the resulting models were
significantly hotter than the best fitting models for the optical
spectra.
However, results from previous (mostly optical) studies showed that
AMES-Cond models are appropriate only for effective
temperatures below about 1700~K, whereas our
fitting process produced significantly hotter
effective temperatures.
Therefore, we also tried to fit the AMES-Dusty models
to the K band spectra, but the resulting temperatures
were about the same and the quality of the fits did not
improve significantly.
As a visual inspection showed, the shape of the K band AMES-Cond spectra
is not very temperature sensitive and mostly dominated by CO
and H$_2$O absorption for models with \teff$\ge$1800~K. 
Below 1800~K, CH$_4$ starts showing in the synthetic
spectra which is further investigated in section \ref{methane}.

\subsection{Absolute flux comparisons}
To clarify on the different effective temperatures we calculated absolute flux values for both synthetic spectra and synthetic magnitudes and
compared them with observed data.
The calculation of synthetic absolute flux values hinges on the knowledge of the radius of the
object.
However, according to \citet{chabrier2000}, the radius of a very low mass object is
0.1$\pm$0.03~R\subsun\ for an age older than 300Myrs and a mass between 0.01~M\subsun\ and 0.07~M\subsun.
These are very conservative limits which are easily fulfilled for a random field sample like the presented 2MASS sample.
Therefore, we used a radius of 0.1~R\subsun\ ($\approx$ 1~R$_{\rm Jupiter}$) to calculate absolute
flux values.

Such a large error in radius will introduce an even larger error in flux.
However, the reason for doing this comparison is to compare flux values
from the ``cool'' (1400K) models and from the ``warm'' (1700K) models.
At these low effective temperatures the flux differences from the effective temperatures are much larger than the
flux differences from the radius uncertainty. Therefore, we can use
the absolute flux values to decide between the results for the effective temperatures
from the optical fits and from the infrared fits.

Table \ref{lowresfittab} lists the observed K$_S$ magnitudes and
K$_S$ magnitudes calculated from the measured trigonometric parallax
the synthetic spectra with the best fitting parameters and the
2MASS K$_S$ response function.
As can be seen, using the parameters from
the optical low resolution spectra fitting give excellent agreement with observations.
The synthetic magnitudes using the results from the
infrared analysis are significantly brighter than the observed magnitudes.

The K$_S$ magnitude of 2MASSs J0850359+105716 is within the error bars
of both the 1700~K model and the 1900~K model synthetic magnitudes
and cannot be used to distinguish between the warm and cold models.
For the two L8 dwarfs with parallaxes it would alternatively require
an assumed radius of about 0.15~R\subsun\ to favour the
synthetic magnitudes from the infrared fitting over the synthetic magnitudes
from the optical fitting.
However, according to \citet{chabrier2000} this would correspond to
objects younger than 50~Myrs and less massive than 0.02~M\subsun\ which have
matching effective temperatures. This is extremely unlikely for a field
sample like this one and would be in contradiction to recent age estimates
for 2MASSW J1523226+301456 (Gl584C) of at least 1Gyr \citep{kirk2001comp}.

\begin{figure}
\plottwo{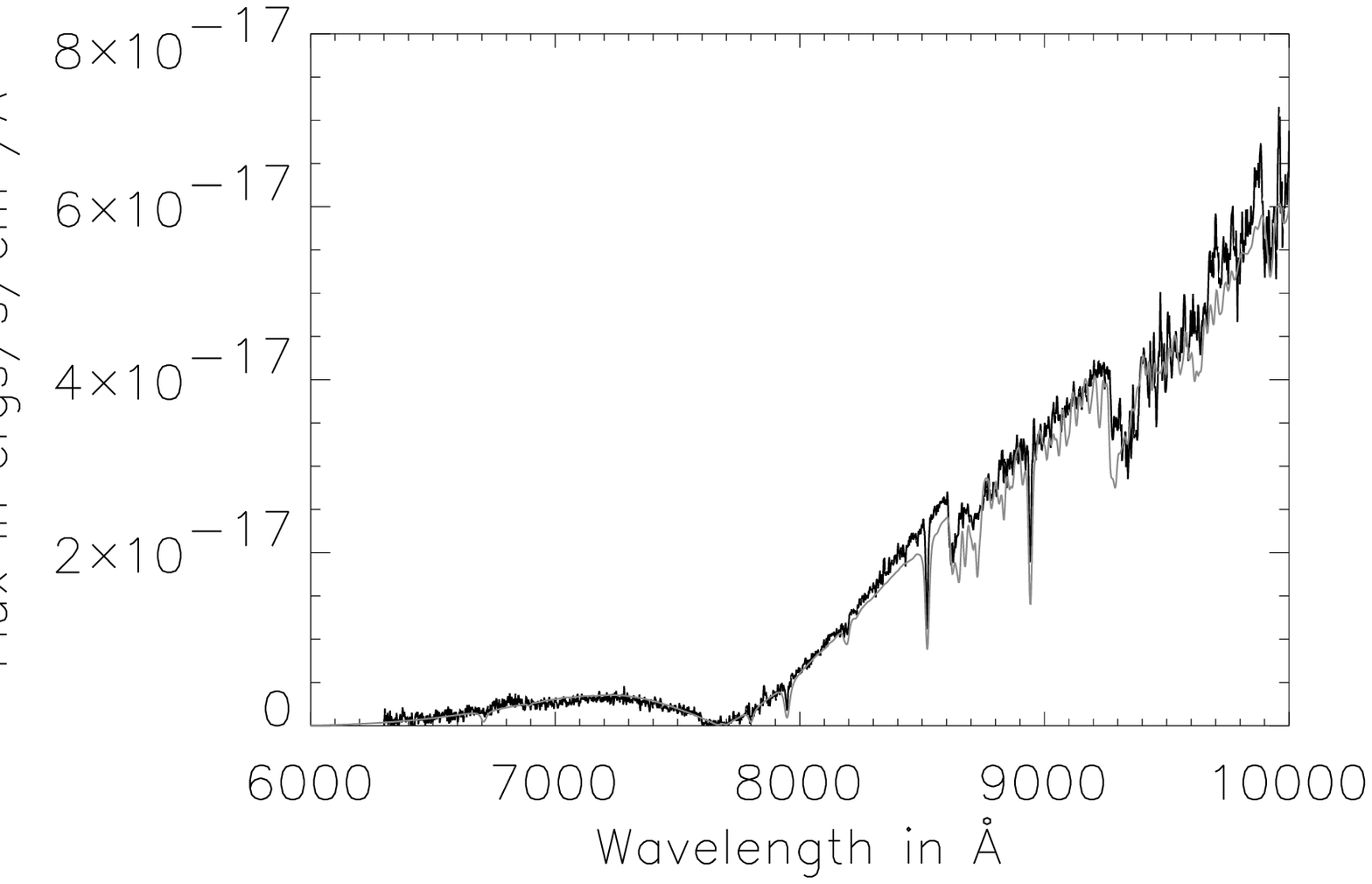}
{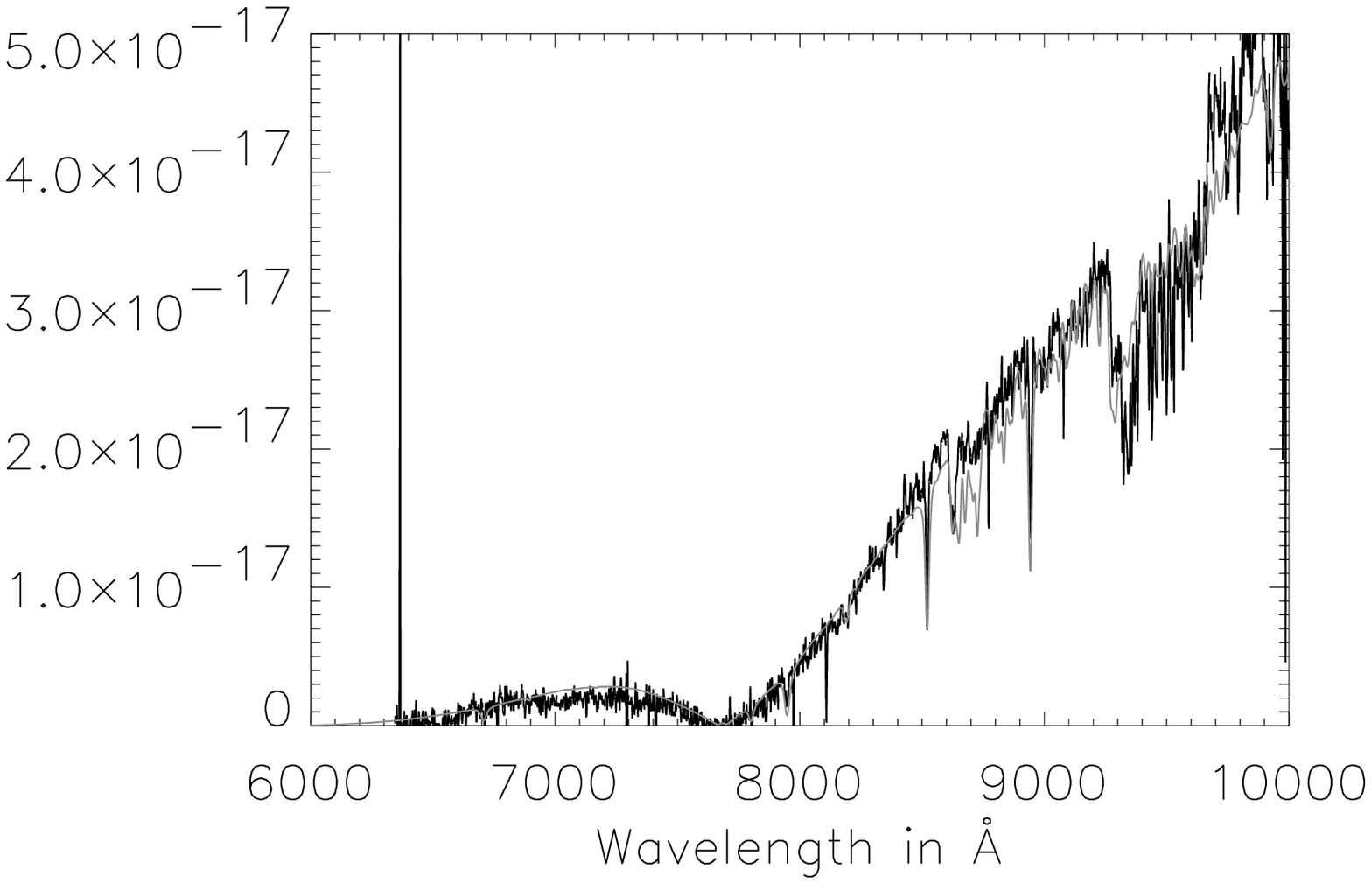}
\caption{\label{absfig}
Models (grey line) and observed spectra (dark line) for
 2M1632+190 (left)  and 2M1523+301 (right) in absolute flux
units.
The only scaling applied to the models are the parallax
and a radius of 0.1~R\subsun.
}
\end{figure}

Figure \ref{absfig} shows 2MASSW J1632291+190441 and 2MASSW J1523226+301456
and the respective best fitting models in astrophysical flux.
The only scaling applied to the synthetic spectra are the distance
and an assumed radius of 0.1~R\subsun. As can be seen, the flux agreement
is excellent.
2MASSs J0850359+105716 is not shown since
the best fitting model (AMES-Cond with \teff=1700K, \logg=5.5)
scaled to the distance and 0.1~R\subsun\ has almost a factor
two more flux than the observation. This is within the expected error range
due to the radius uncertainty. Also, \teff=1700K
is already close to the parameter region between the valid regimes
of the AMES-Cond and AMES-Dusty models. In particular,
AMES-Cond models (when leaving their applicable parameter range) have too much
flux in the optical due to the lack of dust opacity.
Again, using spectra with the parameters obtained
from the infrared fitting results would be brighter by more 
than a factor of two (aside from the different spectral shape).

Since the infrared spectra are not absolute flux calibrated
and since there are no optical magnitudes available we
could not do a corresponding analysis for those quantities.

\subsection{Methane behavior}
\label{methane}
As mentioned above, the AMES-Cond models start showing
CH$_4$ at 2.2\micron\ for \teff$\le$1800~K.
This is the main reason why the K band fitting only yields
effective temperatures above 1800~K.
The AMES-Dusty models do not show CH$_4$ at 2.2\micron;
not even at \teff=1500~K. Below that effective temperature, the AMES-Dusty
models are no longer appropriate \cite[see e.g][]{lhires,cond-dusty}
and their spectra are unrealistic.

However, late L~dwarfs (\teff$\approx$1500~K according to our
optical and flux analysis) do not show CH$_4$ at 2.2\micron.
Yet, \citet{noll2000} have recently reported the detection
of CH$_4$ at 3.3\micron\ for late L~dwarfs.

\begin{figure}
\plottwo{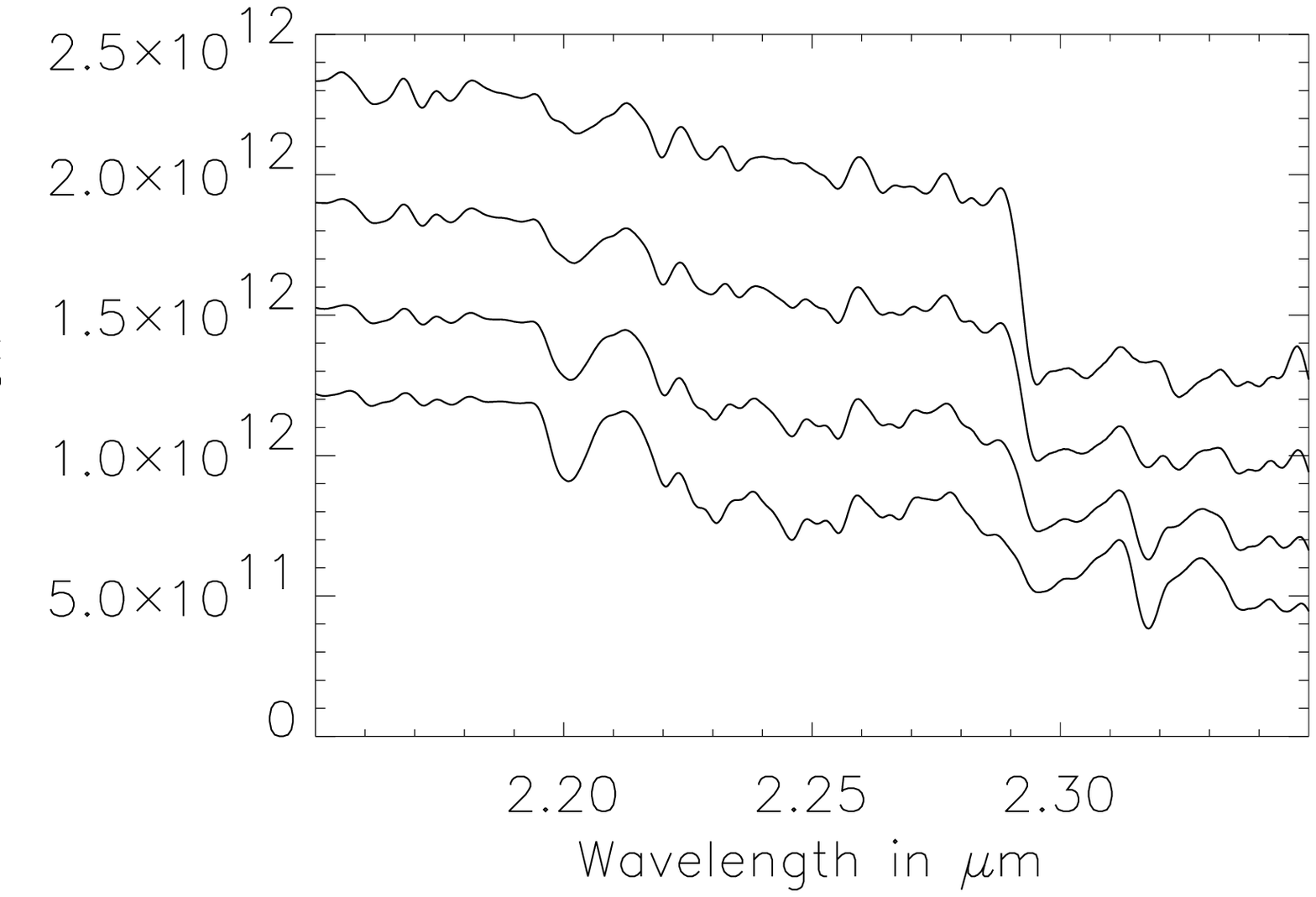}
{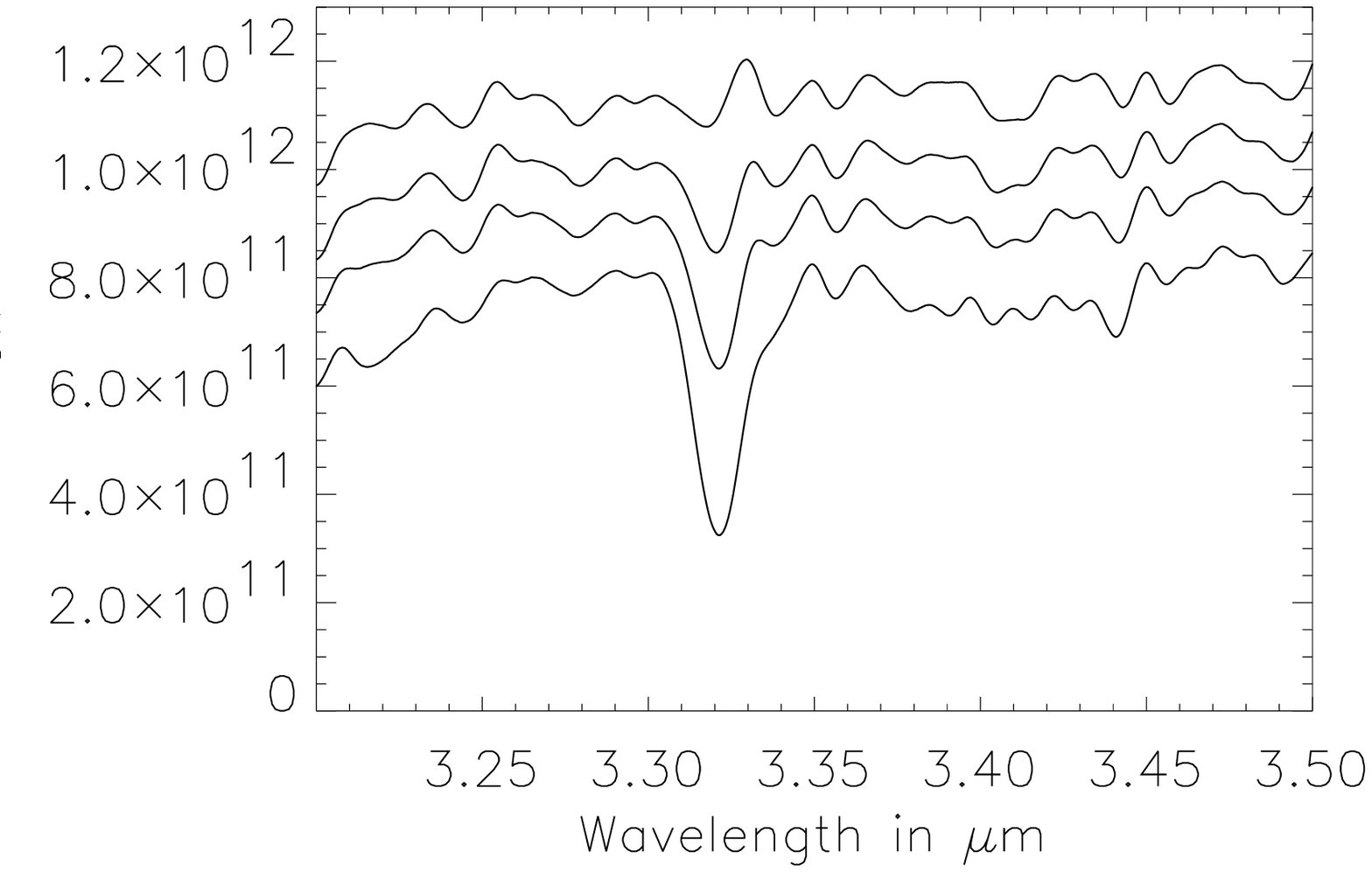}
\caption{\label{ch4fig}
The onset of Methane absorption at 2.2\micron\ (left) and at 3.3\micron\ (right).
All models are for \logg=5.5. \teff is from top to bottom 1900K, 1800K, 1700K, 1600K (left) and
2300K, 2200K, 2100K, 2000K (right). The models have been convolved to yield a resolution of
R=1000.
}
\end{figure}

Therefore, we also investigated  the 3.3\micron\ spectral
region and found the onset of CH$_4$ to occur
at an effective temperature of about 2200~K.
This is about 400~K hotter than the onset of the 2.2\micron\
band. The absorption onset at 3.3\micron\ as well
the absorption onset at 2.2\micron\ are demonstrated in Fig. \ref{ch4fig}.
As can be seen, the CH$_4$ feature is visible at 2.2\micron\ for \teff$\le$1800K
and at 3.3\micron\ for \teff$\le$2200K.
The figure shows models with \logg=5.5, however, the result is the same
for 4.5$\le$\logg$\le$6.0.
Note, that our models are fully self consistent.
Unlike the study in \citet{noll2000} our spectra are a direct
output from the model calculations.

The previous sections suggest that there is a systematic
offset in the theoretical K band (and most likely other infrared band pass) spectra.
They seem to show a typical late L~dwarf shape at about
300-400~K too high temperatures. Therefore,
the CH$_4$ onset temperature of 2200~K at 3.3\micron\ has
to be taken with care. We are much more confident 
in the differential result than the absolute result.
I.e. we are confident that the CH$_4$ absorption at 3.3\micron\ occurs
at about 400~K higher effective temperatures than the  CH$_4$ absorption at 2.2\micron.
This is consistent with the 3.3\micron\ band being a
fundamental band and the 2.2\micron\ band being an overtone band
which only appears when sufficient amounts of CH$_4$ have
formed, i.e. at cooler temperatures.

\section{Summary and discussion}
\label{summary}

We have performed a multi-wavelength analysis of late L~dwarfs.
From optical low resolution fits we determined effective temperatures for
L6 to L8 dwarfs between 1700~K and 1400~K. From K band spectral
analysis we determined effective temperatures between
1900~K and 1700~K.
Comparing absolute flux values (K band magnitudes and absolute flux calibrated
optical spectra), we find excellent agreement with the cooler
effective temperatures.
The mismatch for the K band spectra means that the shape looks
too hot, but the flux is correct.
Most prominently, the AMES-Cond models predict CH$_4$ absorption
at 2.2\micron\ which is not observed.

This can be caused by several factors or a combination of them.
For one, the dust is treated only in a limiting case. A more correct
treatment which accounts for gravitational settling will change the
temperature structure and opacity structure of the atmosphere and hence
parts of the spectrum.
As already discussed in \citet{cond-dusty} and \citet{lhires}, the 
treatment of dust has
a significant impact on the temperature structure.
This together with the improved dust treatment
will also change how the elements are distributed over the
compounds and this results in different concentrations of
CO and CH$_4$ (the most prominent opacities in the K band).
Secondly, there are known
uncertainties in the H$_2$O line list \citep{allard2000}.
The current water opacity distributes the flux in the J to K bandpasses
incorrectly.

This has interesting consequences on the CH$_4$ bands.
We found that the onset of the band at 3.3\micron\ appears at about 400~K higher
temperatures
than the onset of the 2.2\micron\ band.
The numbers for the respective effective temperatures have been
derived from the AMES-Cond models.
Previous studies \citep{lhires,lsandy} have only confirmed their
validity for {\em optical} spectra of late L~dwarfs.
It appears from this study that the K band shape and the onset of
CH$_4$ seems to be reproduced at too high temperatures.
As already discussed above, this sort of effect is consistent with an unrealistic
temperature structure which necessarily arises from
treating dust only in a limiting case.

The AMES-Dusty models seemed to reproduce realistic
spectra in the optical and near infrared when applied to
appropriate objects\citep{lsandy,lhires}.
However, as noted in this study, they do not predict any CH$_4$ absorption.
The future models we are currently working on (Allard et al., in preparation)
will treat the dust formation and settling in a consistent fashion.
These models will then have more realistic temperature structures and
reproduce the spectral features more accurately.
On the other hand, the excellent fits of the AMES-Cond
models in the optical already suggest
that the outer atmosphere is dust free.

From the range of previous estimates on the effective temperatures of
late L~dwarfs we derived relatively low effective temperatures.
This leaves little room for objects between classes L and T.
Although the optical spectra look significantly different, our study found 
strong theoretical evidence that CH$_4$ absorption already takes
place at late L type supporting the results from \citet{noll2000}.
More high signal to noise spectra of L dwarfs in the L band and K band
are needed to underpin these recent results.

\acknowledgments
We are
grateful to Richard Freedman (NASA-Ames) who generously provided VO and
CrH line lists for use in the current 
models.
AS acknowledges support from NASA ATP grant NAG 5-8425 to
the University of Georgia.
JEG and JDK acknowledge the
support of the Jet Propulsion Laboratory, California
Institute of Technology, which is operated under contract
with NASA.
PHH acknowledges support in  part from NASA
ATP grant  NAG 5-3018 and LTSA  grant NAG 5-3619 to  the University of 
Georgia and
partial support from the P\^ole Scientifique de
Mod\'elisation Num\'erique at ENS-Lyon.
FA  acknowledges support from CNRS.
This work was also supported
in part by NSF grants AST-9417242, AST-9731450, and NASA grant
NAG5-3505 and an IBM SUR grant to the University of Oklahoma. Some of 
the calculations presented in this paper were performed on the IBM SP
and the SGI Origin 2000 of the UGA UCNS, on the IBM SP of the San
Diego Supercomputer Center (SDSC, with support from the National
Science Foundation), on the Cray T3E and the IBM SP of the NERSC with support from
the DoE, on the IBM SP2 of the French Centre National Universitaire
Sud de Calcul (CNUSC).  We thank all these institutions for a generous
allocation of computer time.

\newpage 

\begin{deluxetable}{llllrllrll}
\tabletypesize{\scriptsize}
\rotate
\tablecaption{Known and derived parameters for the objects in the sample. \label{lowresfittab}}
\tablewidth{0pt}
\tablehead{
\colhead{} & \colhead{} & \colhead{} & \colhead{} & \multicolumn{3}{c}{Optical fitting}  & \multicolumn{3}{c}{K$_S$ band fitting} \\
\colhead{Name} & \colhead{Spectral Type} & \colhead{K$_S$} & \colhead{$\pi_{\rm trig}$}\tablenotemark{a} & \colhead{\teff}   & \colhead{\logg} & \colhead{K$_{S,\rm Syn}$} & \colhead{\teff}   & \colhead{\logg} & \colhead{K$_{S,\rm Syn}$}}
\startdata
2MASSs J0850359+105716  &  L6   & 14.46$\pm$0.07 & 30.6 & 1700  &  5.5  & 14.49 & 1900 & 6.0 & 14.09 \\
2MASSW J0920122+351742  &  L6.5 & 13.93$\pm$0.08 &      & 1700  &  6.0  &       & 1900 & 6.0 &       \\
2MASSI J0825196+211552  &  L7.5 & 13.05$\pm$0.04 &      & 1700  &  6.0  &       & 1800 & 6.0 &       \\
2MASSW J0929336+342952  &  L8   & 14.62$\pm$0.11 &      & 1400  &  5.0  &       & 1800 & 6.0 &       \\
2MASSI J0328426+230205  &  L8   & 14.84$\pm$0.13 &      & 1400  &  6.0  &       & 1700 & 6.0 &       \\
2MASSW J1632291+190441  &  L8   & 13.98$\pm$0.05 & 16.8 & 1400  &  5.5  & 14.06 & 1700 & 6.0 & 13.28 \\
2MASSW J1523226+301456  \tablenotemark{b}
                        &  L8   & 14.24$\pm$0.07 & 18.8 & 1400  &  5.5  & 14.30 & 1800 & 6.0 & 13.26 \\
\enddata

\tablenotetext{a}{from Kirkpatrick et al. 2000}
\tablenotetext{b}{Gl 584C}
\tablenotetext{c}{The error bar on the synthetic magnitudes is about 0.5 magnitudes from the radius uncertainty of 0.03~R\subsun\ (see text for details)}

\end{deluxetable}

\newpage

\clearpage

\clearpage

\end{document}